\documentclass[acus]{JAC2003}

\usepackage{graphicx}
\usepackage{booktabs}

\begin{document}
\title{Yoctosecond photon pulse generation in heavy ion collisions}

\author{A. Ipp\thanks{ipp@hep.itp.tuwien.ac.at}, Vienna University of Technology, Austria}

\maketitle

\begin{abstract}
Heavy ion collisions at RHIC and at the LHC can
create the quark-gluon plasma, a state of matter at very high
temperatures. Among a plethora of particles that are
produced in these collisions, also light is emitted
throughout the evolution of the plasma.

In this talk, the properties of this light are discussed
and related to recent efforts towards shorter and more
energetic photon pulses in laser physics. In particular,
the time evolution of high-energy photons is studied.
These photons originate from Compton scattering of gluons and
quark-antiquark annihilation in the plasma. Due to the internal
dynamics of the plasma, double pulses at the yoctosecond time
scale can be generated under certain conditions. Such double
pulses may be utilized for novel pump-probe experiments
at nuclear time scales.
\end{abstract}

\section{INTRODUCTION}
The year 2010 marks the fiftieth anniversary of the invention of the laser.
Since its invention, not only did the intensity steadily increase,
but also the pulse duration became shorter and shorter.
In fact, pulse duration and intensity of lasers
(or derived coherent radiation bursts) turn out
to be correlated over a large range of energies and
time scales \cite{Mourou:2011}.
This observation provides a good motivation to 
present at a conference on Physics in Intense Fields (PIF 2010)
a study of the shortest possible light flashes
that can be produced in experiments,
which is heavy ion collisions that produce the quark-gluon plasma (QGP)
\cite{Ipp:2009ja}.

\begin{figure}[htb]
   \centering
   \includegraphics*[width=65mm]{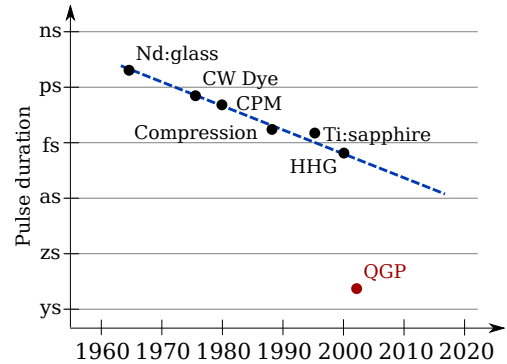}
   \caption{
History of laser pulse duration. Marked are selected milestones
of technologies that allowed to decrease the laser pulse duration,
like 
Neodymium glass laser (Nd:glass), 
Continuous Wave Dye laser (CW Dye), 
Colliding Pulse-Mode locked dye laser (CPM),
Titanium sapphire laser (Ti:sapphire), or
High-Harmonic Generation (HHG).
For comparison, also the lifetime of the Quark Gluon Plasma (QGP) is
indicated as it is produced nowadays in heavy ion colliders like RHIC or LHC.
}
\label{laserhistory}
\end{figure}

A few selected milestones of the development of short laser pulses are depicted in Fig.~\ref{laserhistory}.
Each of the time scales has enabled to access new systems:
At the picosecond to femtosecond timescale, femtochemistry allows for
the time-resolved study of chemical reactions~\cite{zewail:1994}.
For pump-probe spectroscopy, it is essential to have two short pulses
in close succession: The first laser pulse triggers a chemical reaction,
which may involve an excited state and various short-lived transitions,
while the second pulse takes a snapshot of the intermediate state.
By varying the interval between the two pulses, the time-evolution
of a chemical reaction can be studied.

At shorter timescales, attosecond science is the window to capturing electron motion
in molecules and atoms~\cite{Drescher:2001}.
High-order harmonics of femtosecond laser radiation have been shown to be sources of trains of attosecond 
extreme-ultraviolet pulses~\cite{Paul2001} that can be used to produce single attosecond
soft
X-ray~pulses~\cite{Silberberg2001}. For example, such single attosecond X-ray bursts~\cite{baeva07} have applications in molecular imaging~\cite{Lein2005},
quantum control~\cite{Rabitz2000}, or Raman spectroscopy~\cite{Dudovich2002}. By introducing a controlled delay between two such peaks, the dynamics of electron systems could be studied using pump-probe techniques \cite{Silberberg2001}. 
Such techniques also allow for the direct time resolution of 
many-body dynamics, like the observation of
the dressing process of charged particles~\cite{Huber:2001}.
In the zeptosecond regime, nuclear processes may become accessible~\cite{Golabek:2009}.
It has been suggested that zeptosecond pulses could be created 
via nonlinear Thomson backscattering~\cite{lan:066501,Kim}, or by employing relativistic
laser-plasma interactions~\cite{Bulanov,Nomura},
A possible detection scheme for the
characterization of short $\gamma$-ray pulses of MeV to GeV energy
photons down to the zeptosecond scale has been proposed in Ref.~\cite{Ipp:2010vk}.

At even shorter timescales, double pulses of yoctosecond duration
of GeV photon energy could be created from the quark-gluon plasma in non-central heavy ion
collisions~\cite{Ipp:2009ja}. 
It turns out that the emission envelope can depend strongly on the internal dynamics of the QGP.  Under certain conditions, a double peak structure in the emission envelope may be observed. This could be the first source for pump-probe experiments at the yoctosecond timescale. The delay between the peaks is directly related to the isotropization time, and the relative height between the peaks can be shaped by varying photon energy and emission angle.  Such pulses could be utilized, for example, to resolve dynamics on the nuclear timescale such as that of baryon resonances~\cite{Dugger:2007bt}. Conversely, a time-resolved study of the emitted photons could provide a window to the internal QGP dynamics throughout its expansion.

\section{QGP PHOTON PRODUCTION}

The Large Hadron Collider (LHC) at CERN has begun its heavy-ion program in November 2010~\cite{Aamodt:2010pa}, just a few weeks before the PIF 2010 conference. 
With center-of-mass energies of 2.76 TeV per nucleon, the collisions of lead ions produce hotter and denser plasmas than previously achievable at the Relativistic Heavy Ion Collider (RHIC), where gold ions were used.
The temperatures reached in these collisions are so high that the constituents of atomic nuclei, the neutrons and protons, are split into their constituents, the quarks and gluons. 
The interest in the QGP stems not least from the fact that it is believed to have filled the entire universe during the first few microseconds after the Big Bang. 

\begin{figure}[htb]
   \centering
   \includegraphics*[width=65mm]{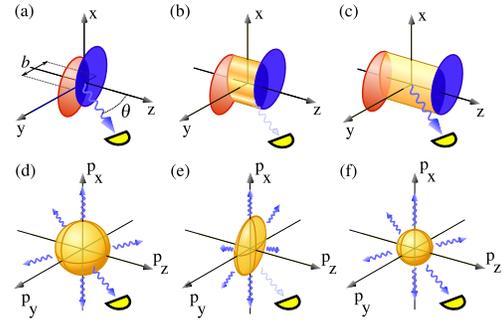}
   \caption{Early stages of a high-energy collision, involving pre-equilibrium (first two columns) and equilibrated QGP phases (last column).
Parts (a)-(c) show three snapshots in time in position space. Shown are the two relativistically contracted colliding ions that create the quark-gluon plasma in the overlap region. Curly arrows denote photon emission and semicircles the detectors. Parts (d)-(f) are corresponding pictorial representations of the plasma in momentum space. In an intermediate stage
of the pre-equilibrium phase,
the momentum distribution is anisotropic, resulting in a change in the angular photon emission pattern that can give rise to double-peaked photon pulses~\cite{Ipp:2009ja}.}
  \label{fig:system}
\end{figure}

In heavy-ion collisions, the QGP is produced up to the size of a nucleus ($\sim15\,\mbox{fm}$) for a duration of a few tens of yoctoseconds ($1\,\mbox{ys}=10^{-24}\,\mbox{s}$). The plasma is produced initially in a very anisotropic state, and reaches a hydrodynamic evolution through internal interactions only after some isotropization time $\tau_{\rm iso}$ (see Fig.~\ref{fig:system}). 
Among the many particles that are produced, also photons of a few GeV energy are emitted~\cite{Fries:2002kt,Ipp:2007ng}. 

It was one of the early surprises that the observed particle spectra turned out to agree well with ideal hydrodynamical model predictions~\cite{Huovinen:2001cy}. 
This led to the initial assumption that 
isotropization times may be as low as $\tau_{\rm iso}\approx1\,\mbox{ys}$.
However, it has been pointed out in the mean-time that viscous hydrodynamic models are still consistent with RHIC data if isotropization times as large as $\tau_{\rm {iso}}\approx7\,\mbox{ys}$ are assumed \cite{Luzum:2008cw}, even if the expansion before isotropization is assumed to be collisionless ({}``free streaming''). 
This should be compared to the lifetime of the QGP, which could amount to 15 ys at RHIC, and which could be as large as 25 ys at LHC.

Figure \ref{fig:system}(a-c) shows a schematic view of the collision of two heavy ions.
The ions are illustrated as relativistically contracted pancakes.
In general, two ions will not collide head-on-head, but will be displaced by an impact parameter $b$.
Direct photons are emitted from the expanding QGP throughout its lifetime~\cite{Turbide:2003si}.  The energy spectrum of the emitted photons extends to the GeV range, and the upper limit for the temporal duration of the GeV photon pulse is determined by the expansion dynamics of the QGP, which leads to yoctosecond pulses. 
At the initial stage of the collision, even before the plasma is created, prompt photons are emitted from nucleon-nucleon collisions in all directions, see Figs.~\ref{fig:system}(a) and (d).
For an intermediate time after the collision shown in Fig.\ref{fig:system}(b), a momentum anisotropy occurs 
due to the longitudinal expansion of the plasma:
Those particles that originally had momentum components in forward or backward direction along the beam axis leave the central region quickly, so that mainly particles with transverse momenta remain in the plasma.  
High-energy photons that are created in the plasma through Compton scattering or quark-antiquark annihilation carry preferentially the momentum of the original participants of the collision.
Because of the momentum anisotropy of the quarks and gluons within the plasma, also the emitted photons are preferentially emitted perpendicular to the beam axis $z$, as indicated in Fig.~\ref{fig:system}(e).
Finally, in Fig.~\ref{fig:system}(c), the system had time to isotropize due to collisions within the plasma. The photons will be emitted again in all directions, as shown in Fig.~\ref{fig:system}(f), particularly also 
into the direction in which the photon emission was suppressed during the anisotropic stage.

A photon detector placed towards the beam axis would therefore measure a time-dependent photon flux. Ideally, in the intermediate stage in Fig.~\ref{fig:system}(e), the photon emission in the direction of the detector is suppressed highly enough, so that the photons emitted in the stages Fig.~\ref{fig:system}(d) and Fig.~\ref{fig:system}(f) are distinct enough in time to form two separate photon pulses. 
In order to quantitatively describe the pulse envelope of the emitted photons, detailed calculations are necessary.
Such a calculation has been performed in Ref.~\cite{Ipp:2009ja}, which was based on a one-dimensional expansion model by Bjorken \cite{Bjorken:1982qr}. This model assumes boost-invariant evolution of the quark-gluon plasma in a central region of the collision. The detector is placed away from the beam axis by an angle $\theta$
within the reaction plane.
Direct photons in the GeV energy range are 
emitted from the pre-equilibrium and
equilibrated phases of the QGP. 
The leading contribution to the photon production rate $R$ originates from quark-gluon Compton scattering and quark-antiquark annihilation.
In principle, higher order soft scattering
processes like bremsstrahlung or inelastic pair annihilation would have to be included as well,
but their contributions become dominant only at lower energies, and are less important at the higher energies considered~\cite{Schenke:2006yp}.

For anisotropic momentum distributions, the photon production rate $R$ 
has to be calculated numerically~\cite{Ipp:2007ng}.
It depends on the temperature $T$ of the medium, the photon energy $E$ and momentum $k$, the fine structure constant $\alpha$, and the corresponding quantity for the strong force $\alpha_{s}$ (with $\hbar=c=k_{\rm B}=1$).
The rate further depends on the anisotropy, which is described by a parameter 
$\xi=\left\langle p_{T}^{2}\right\rangle /\left(2\left\langle p_{L}^{2}\right\rangle \right)-1$
that relates the mean longitudinal and transverse momenta 
$p_{L}$ and $p_{T}$~\cite{Schenke:2006yp,Mauricio:2007vz}.
To integrate this rate over time, a time evolution model for the pre-equilibrium and equilibrated QGP has been used~\cite{Mauricio:2007vz}.
This model specifies the time evolution for the energy density $\mathcal{E}=\mathcal{E}(\tau)$, for the hard scale $p_{{\rm hard}}=p_{\rm hard}(\tau)$ (which corresponds to $T$ in the isotropic case), and for the anisotropy
parameter $\xi=\xi(\tau)$ as a function of the proper time $\tau$. 
Qualitatively, the model follows the evolution as outlined in Fig.~\ref{fig:system}. For early times, a free streaming phase lets the anisotropy grow. At late times, the system converges to an ideal hydrodynamic phase with vanishing anisotropy. These two phases are linked by a smooth transition which is controlled by additional model parameters. Thermalization and isotropization happen concurrently in this model, $\tau_{\rm {therm}}=\tau_{\rm {iso}}$.
The model is thus able to cover both, the pre-equilibrium phase and the equilibrated QGP phase of
the expanding plasma.

\begin{figure}[htb]
    \centering
    \includegraphics*[width=65mm]{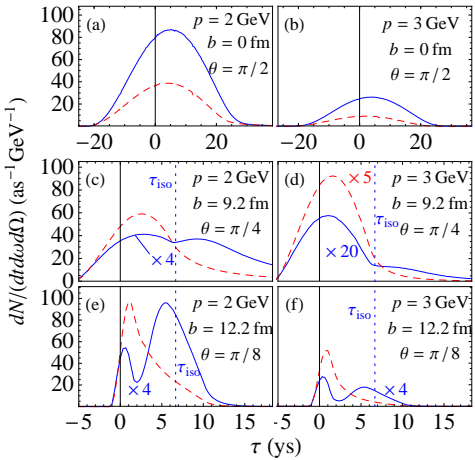}
    \caption{Photon emission rate as a function of detector time $\tau$.
Solid blue lines show a large isotropization time $\tau_{\rm {iso}}=6.7\,\mbox{ys }$ while dashed red lines correspond to a short isotropization time $\tau_{\rm {iso}}=0.3\,\mbox{ys}$. 
Parts (a) and (b) display emission at midrapidity ($\theta=\pi/2$) for a central collision with impact parameter $b=0$.
Parts (c)-(f) show double-peaked 
photon pulses obtained for $b=9.2$~fm or $12.2$~fm, and the vertical dotted line indicates the position of the larger $\tau_{\rm {iso}}=6.7$~ys. 
In parts (c) and (d), the detector direction is $\theta=\pi/4$, in (e) and (f), it is $\theta=\pi/8$~\cite{Ipp:2009ja}.}
  \label{fig:result}
\end{figure}

The numerical parameters suitable for calculating LHC parameters are given as follows: 
The initial temperature is assumed to be $T_{0}=845$ MeV with a formation time $\tau_{0}=0.3\,\mbox{ys}$.
The critical temperature, where the QGP ceases to exist, is taken as $T_{C}=160$ MeV. 
The isotropization time is varied in the range $\tau_{{\rm {iso}}}=\tau_{0}$ to $\tau_{{\rm {iso}}}=6.7\,\mbox{ys}$, assuming free-streaming at early times.
Both possibilities are not yet ruled out by RHIC data. In order to ensure fixed final multiplicity, the initial conditions are adjusted as a function of $\tau_{\rm {iso}}$ such that the same entropy is generated as for $\tau_{\rm {iso}}=\tau_{0}$.

For central collisions with emission angle orthogonal to the beam axis ($\theta=\pi/2$), a typical time evolution of the photon emission rate is depicted in Figs.~\ref{fig:result}(a) and \ref{fig:result}(b).
At 2 GeV energy, the photon production from the QGP at midrapidity
is 3 to 4 times as large as the production from the initial collisions.
It is roughly 6 times as large as the 
production from the hadron gas~\cite{Turbide:2003si}.
At 3 GeV energy, QGP photon production is even 50 times larger than the production from the hadron gas.
In the Figs.~\ref{fig:result}, the origin of the abscissa is the time when a photon emitted from the center of the collision arrives at the detector. Photons arriving earlier originate from a part of the QGP that is closer to the detector. The pulse shape is mainly determined by the geometry of the lead ion with radius $7.1\,\mbox{fm}$. Any structure on the yoctosecond timescale is blurred simply by the time for light to traverse the QGP. 

This limit can be overcome in the following ways: By considering non-central collisions with impact parameter $b$, the physical extent of the QGP is reduced. Also, an optimization of the detection angle can minimize the traveling time through the plasma. In forward direction, the initial shape of the QGP is Lorentz-contracted, and light leaves this initial region quickly. This is partially spoiled due to the QGP expansion in the same direction. Thus intermediate emission angles are most promising for which the QGP appears partly Lorentz contracted but does not expand towards the detector.
                                                                                                                                                                                    
Figures~\ref{fig:result}(c)-\ref{fig:result}(f) show the photon emission in the directions $\theta=\pi/4$ and $\theta=\pi/8$. For large impact parameters $b=9.2$~fm and $b=12.2$~fm, a striking double-peak structure appears. The minimum between the two peaks corresponds roughly to maximum anisotropy within the plasma. This follows from the fact that the photon emission
rate is suppressed for larger values of $\xi$ and smaller values of $\theta$~\cite{Schenke:2006yp}. The distance between the two peaks is approximately governed by the isotropization time $\tau_{\rm {iso}}$, indicated by the dotted line in Figs.~\ref{fig:result}(c)-\ref{fig:result}(f). 
The first peak corresponds to photons emitted from the blue-shifted approaching part of the QGP, 
while the second peak corresponds to 
a slightly red-shifted and time-dilated receding tail of the plasma.
For a short isotropization time $\tau_{\rm {iso}}=\tau_{0}$ (dashed lines) the separation into two peaks does not occur. Therefore, this effect depends delicately on the QGP dynamics.

There are a couple of caveats to this model calculation:
Apart from the photons originating from the QGP, in an actual experiment there is a background of photons from various sources~\cite{Turbide:2005fk}. These include photons produced by a jet passing through the QGP~\cite{Fries:2002kt}, and could dominate the effect that is expected from the QGP alone. 
Since these photons are produced on a similar yoctosecond timescale, they would modify the pulse shape on this timescale.
Photons produced from the initial collisions of the two ions can be of comparable size in intensity,
but they would only enhance the first peak of the double peaks depicted in Fig.~\ref{fig:result}(c)-\ref{fig:result}(f).
Other background photons are produced in the decay of pions at later stages of the collision.
Since these are produced at much later time scales, they would not modify a time structure on the yoctosecond timescale.
It would be necessary to take these various photon sources into account in order to obtain a quantitative prediction of the expected pulse structure.

What kind of properties should the detector have in order to resolve the double pulses?
The time-integrated high-energetic photons are already routinely detected in experiments \cite{Adler:2005ig}.
In the GeV energy range, the photon production rate is of the order of a few photons per collision~\cite{Turbide:2005fk}. Note that a single GeV photon pulse of 10~ys duration corresponds to a pulse energy of only about $100$~pJ, but to a power of $10$~TW. 
The photon yield could be enhanced by considering lower energy photons,
but this would also 
increase the number of unwanted background photons.
Alternatively, the photon yield could be enhanced by increasing the collision energy.
This may have the additional benefit of increasing the relative importance of the contribution of thermal photons compared to other kinds of photons~\cite{Turbide:2005fk}. 
The emission envelope is influenced by the geometry, emission angle, and internal dynamics like the isotropization time of the expanding QGP. The double-peak structure described may emerge in non-central collisions at an emission angle close to forward direction, assuming that the isotropization time is large.
In order to detect such short pulses, new detection schemes would be required.
Existing tools and ideas from attosecond metrology, like pump-probe experiments or spectroscopy techniques, may turn out to be appropriate candidates to be scaled to zepto- or 
yoctosecond duration~\cite{Drescher:2001,Ipp:2010vk}. Also, an experimental determination of the photon emission envelope would serve as an additional probe of the internal dynamics of the QGP, for example by measuring its isotropization time.

\section{CONCLUSIONS}
The steady progress in laser physics over the last decades towards higher intensities
and shorter pulse lengths provides strong motivation to study systems
that produce extremely short flashes of light.
The QGP is an ideal candidate because it exhibits non-trivial dynamics on the 
yoctosecond timescale, during which GeV photons are emitted.
Under certain conditions, a double-peak structure may be produced.
This could eventually lead to novel pump-probe experiments at the GeV energy scale.
Alternatively, measuring the temporal shape of the photon emission envelope,
dynamic properties of the QGP, like its isotropization time, could be probed experimentally.

\section{ACKNOWLEDGMENT}

I would like to thank my collaborators J.~Evers, K.~Z.~Hatsagortsyan, and C.~H.~Keitel for guidance and fruitful discussions that led to the work that has been presented \cite{Ipp:2009ja,Ipp:2010vk}. I would further like to thank the organizers of the PIF 2010 for their kind hospitality.

\end{document}